# A MODEL FOR SEMANTIC INTEGRATION OF BUSINESS COMPONENTS

Larbi KZAZ[1] , Hicham ELASRI[2] and Abderrahim SEKKAKI[3]

[1]*Higher Institute of Commerce and Business Administration (ISCAE) Km 9,500 Route de Nouasseur P.O Box. 8114 - Casablanca Oasis. Morocco Phone: +212-522-335482/83/84/85 Fax: +212-522- 335496*

1kzaz_larbi@yahoo.fr

[2,3]*University Hassan II Aïn-chok, Faculty of Sciences Department of Mathematics & Computer Science P.O Box 5366, Maarif – Casablanca. Morocco Phone :+212 -522-230684 Fax : +212 -522- 230674*

2hicham_elasri@yahoo.com

[3]a.sekkaki@fsac.ac.ma

### ABSTRACT

***Reusable components are available in several repositories; they are certainly conceived for reusing in developing information system, however, their re-use is not immediate; it requires, in fact, to pass through some essential conceptual operations, among them in particular, research, integration, adaptation, and composition. We are interested in the present work to the problem of semantic integration of heterogeneous Business Components. This problem is often put in syntactical terms, while the real stake is of semantic order. Our contribution concerns a model proposal for Business components integration as well as resolution method of semantic naming conflicts, met during the integration of Business Components.***

### KEYWORDS

*Business Component, Semantic Integration, Ontology, Semantic Web.*

## 1. INTRODUCTION

This work is placed in the context of component based software development approach; more precisely, we are interested to semantic integration of a particular category of components called "*Business Component*" (BC). Several types of semantic conflicts must be resolved during the integration of BC; our work focus on naming conflicts. Our goal is to support BC reuse by providing a model, an integration method based on ontologies, and a similarity measure calculating based method for the resolution of BC naming conflicts. After a review of related works on component based software approach in section 2 we describe the concept of business component paradigm in section 3; in section 4 problems of BC semantic integration and different types of conflicts that raise are listed, then ontologies and their use in the Semantic Web are presented, and a proposal to extend their application to BC is done. In section 5 a model and an ontology-based on methodology for BC semantic integration are presented. Examples of applying the method are also showed. Finally, perspectives of extending this research are outlined and concluded.

## 2. RELATED WORKS

Components based approach is considered since earliest 1990's as a new information system development paradigm [3]. This approach aims to reduce significantly costs and cycle-time of developing software. Components based approach consists in building new systems by reusing available components. Using this approach in the earliest phases of system development presents a real interest [16]. Two ways of research in the area of the reuse are intensively explored. The first one called "design for reuse" is to develop methods and tools to produce reusable components. The second "design by reuse", is to develop methods and tools to exploit reusable components [17]. We are concerned in this research by the second way.
Literature outlines several questions when we address the topic of designing a new Information system by reusing available components. In fact, the reuse of components requires several operations such: research, selection, adaptation, composition [17] and integration. This last operation has been identified by [3], the author also points the axis of semantic integration. Our work focuses on the issue of semantic integration of business components. We rely to address this issue on the results obtained in semantic web and knowledge engineering fields. Ontologies have been intensively used in this field to describe the semantic. We will rely on ontologies to detect semantic conflicts on Business components.

## 3. Business Component Paradigm

The term of component is widely used in the field of reusing, with a general connotation of reusable entity. The aim of Information Systems (IS) development based on component-approach is to construct IS from a set of available reusable components. The Business Component paradigm is based on a particular category of components called Business Components (BC).

### 3.1. Definition

Several definitions of the concept of business component are encountered in literature. We retain two definitions: the first is that given by Sims and Herzum in [10] *"A business component is the software implementation of an autonomous business concept or business process. It consists of all the software artefacts necessary to represent, implement, and deploy a given business concept as an autonomous, reusable element of a larger distributed information system"*. The second is that given by F. Barbier [3] *« A business component models and implements business logic, rules and constraints that are typical, recurrent and comprehensive notions characterizing a domain or business area »*.

According to BC paradigm, a company IS is built from a set of BC which can be originated from multiple sources. The IS commercial company, for example, could be designed from BC such: {«"Sales", "Product", "Customer" etc . . .}

### 3.2. Classification

Classification of business components (BC) may be based on the "type of knowledge" that represents. According to [10] a BC can be an entity: Entity-BC (Employee, customers, suppliers, addresses, invoices, etc..) Or a process: Process-BC (procurement process, sales process, etc...) Or an utility: Utility-BC (Note, Code, etc). To these three categories, [8] added a fourth one: Data-BC. We note that the last two categories: Utility-BC and Data-BC, have low granularity, and are not intended to be used independently from the two other categories components; they serve to them as a basis for their design. On the contrary, Process-BC and Entity-BC, which are of high granularity, can be reused independently. We can deduce from above the existence of some hierarchy among BC. Process-BC which are at the top level, are based on the Entity-BC, these last are located at the next level. Utility-BC are at the level immediately below, followed by Data-BC that are placed at the lowest level of the hierarchy [3].

Another classification distinguishes between "vertical components", only reusable within the same domain, and "horizontal components", which are reusable in different domains [16]. We must also note that it is necessary to distinguish between a conceptual and software aspects of BC concept [13] and [3].

Software components are described in programming languages and for given infrastructures, while conceptual components are described in standard, technologically and neutral modeling language, such as UML. The component-based development can be defined as the reuse and integration of models describing components [21]. In the remainder of this paper, the term BC will design the conceptual level aspect. This aspect is fundamental in the activities of specification and design of IS based on BC approach.

# 4. SEMANTIC INTEGRATION

## 4.1. Definition

Sandra Heiler [8] introduced semantic interoperability term in 1995, she defined it as following: *« Interoperability among components of large-scale distributed systems is the ability to exchange services and data with another. (…) Semantic interoperability ensures that these exchanges make sense – that the requester and the provider have a common understanding of "the meanings" of the requested services and data. »,* [3], [14], [20] [22]. Semantic Interoperability is the ability of components to exchange data and services while sharing their sense. Thus, semantic interoperability enables semantic integration. Therefore, semantic integration of components requires detection and resolution of semantic conflicts that may exist among components.

## 4.2. Integration conflicts

Data and service exchange among BC can give rise to different types of semantic conflicts. Several researchers [9], [12] [11] [23] identified three types of semantic conflicts: confusion, measure and naming conflicts.

### *4.2.1. Confusion conflicts*

[9] has defined confusion conflict as follows: *« Confounding conflicts: information items appear to have the same meaning but differ in reality due to e.g. a different temporal context (e.g. 'occurred 5 minutes ago') »*

Confusion conflict type is therefore linked to contextual data with the same appearances, but changes behavior over time. For example, if an employee worked as manager in the past, he could be a Director today so he's still an employee but he is promoted. The business component «employee» illustrates this case in the figure below

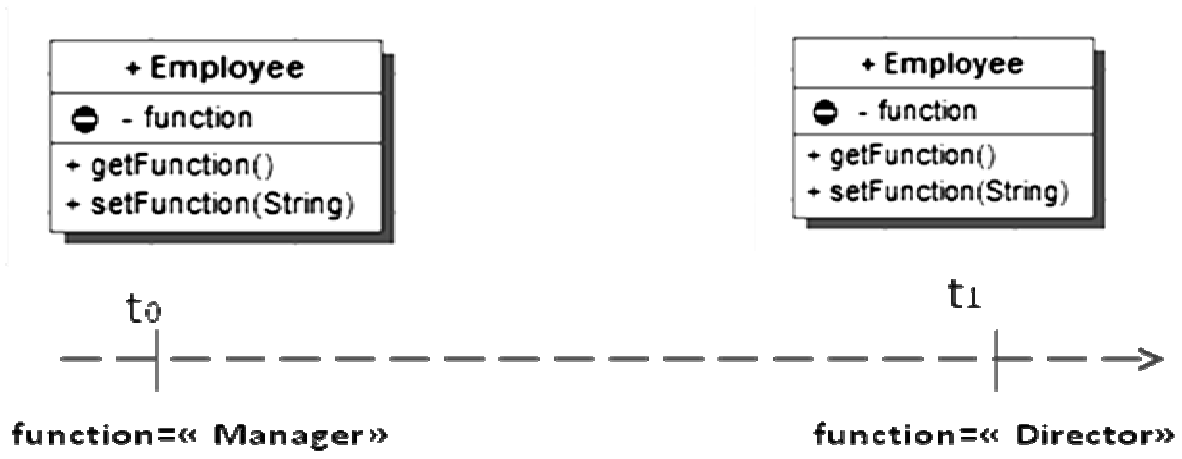

**Figure 1 : Confusion conflict example**

### *4.2.2. Measure Conflict*

Measure conflict occurs when two systems express the same value with different units [9] [11] [23]. For example if we want to integrate two business components "Product", including the attribute "Price". The attribute measure unit of the first component uses the Euro, and the attribute of the second component uses the dollar (see Figure 2).

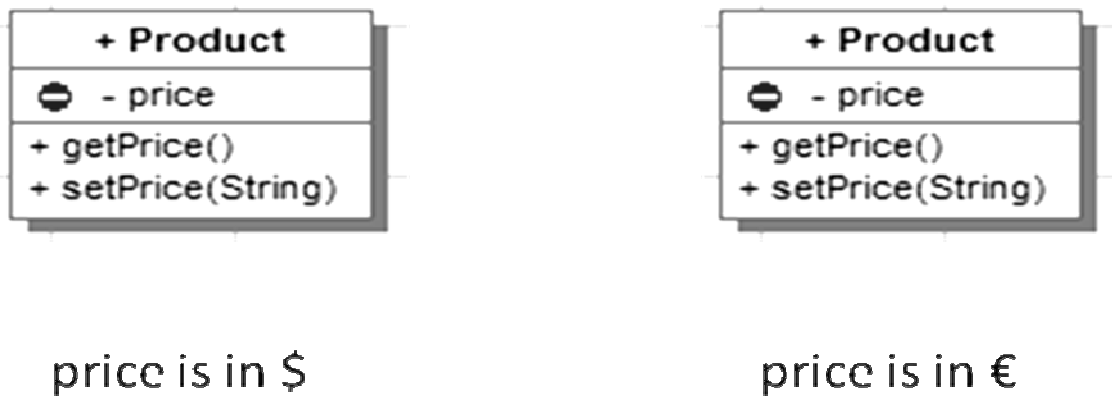

**Figure 2: Measure conflict example**

#### *4.2.3. Naming Conflict*

[23] Defines naming conflict type as follows: *«Naming conflicts occurs when naming schemes of information differ significantly. A frequent phenomenon is the presence of homonyms and synonyms »*. Naming conflicts are due to the presence of homonyms and synonyms. The example below illustrates the synonymy phenomenon among two BC "Client" and "Customer". These two BC are synonyms; they have different names and represent the same component.

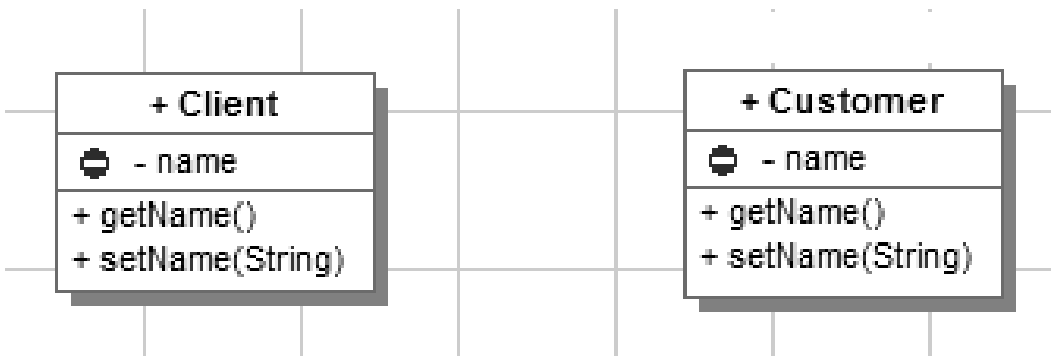

**Figure 3: Example of synonymy phenomenon.**

We will be exclusively interested in the following of this paper to naming conflict.

### 4.3. Integration mechanisms

An integration mechanism aims to resolve conflicts due to BC heterogeneity, in order to make them interoperable. There are two types of integration mechanisms in literature: mechanisms based on predetermined models (component models) and mechanisms based on ontologies. Our proposal will be based on ontologies, considered as a key element to ensure conflict resolution.

#### *4.3.1. Ontologies.*

Ontologies offer a common and shared understanding of a domain, for both human users and software applications. They have become a key tool in knowledge representation, their applications are numerous and subject of intense work in different areas: artificial intelligence, natural language processing, information retrieval, collaborative work etc. [1] According to [17], a domain ontology is defined by two complementary elements:

**The domain model**: It's composed of concepts and relations among these concepts. It includes the concepts contained in the local ontologies (source ontologies) and introduces taxonomy (hierarchical structure) of these concepts.

**The thesaurus:** It contains derived terms and definitions of domain model concepts. It also provides vocabulary for describing generic contexts. Derived terms are synonymous and homonymous concepts. Ontology Alignment (mapping research, matching or mapping) is a particularly important task in systems integration; this topic has given rise to many works [19], we rely on these works results to achieve BC semantic integration. BC represents in our case, the resources to integrate through ontologies.

Our use of domain ontologies can be justified by several reasons: First of all domain ontology concerns, by definition, concepts relating to a particular application domain, this complies perfectly with our problem, since the design of an IS concern usually a business area. Second, domain ontologies are reusable within the same area [4] [15] this property is very interesting in BC vertical reuse, which is the central aim of component-based approach.

#### *4.3.2. The semantic web*

According to the W3C, *« the Semantic Web is a vision the idea is that the data are on the web defined and linked in order to be used by machines on the web not only for display, but for automation, integration and reuse on various platforms»*. Semantic Web is an extension of current Web giving meaning to the content. It leans between others, on ontologies and languages, in order to give and represent meaning of its resources, and so as to allow programs and agents to access it using languages developed by W3C.

In our case, the Semantic Web can be used as a platform for designers who are searching for reusable components in order to design a new information system. In fact, several studies focus on the process of finding reusable components. these aim to simplify the reuse of components by means of more or less natural approaches, which purpose is to provide components that are adapted to the IS designer needs[17].

# 5. A PROPOSAL FOR BUSINESS COMPONENT SEMANTIC INTEGRATION

## 5.1. Business component integration

BC provides both services and data; BC semantic integration aims to assign meaning to data and services to ensure data and services exchanges between heterogeneous BC.

## 5.2. Business component integration model

The integration model that we suggest exploits the results of some works on components and ontologies:

- The transformation to ontologies of BC described in a modeling language like UML. This transformation is made possible relying to [5],[6]. This work presents an approach based on XSLT language for automatic generation of OWL from UML description.
- The alignment of ontologies, obtained from the transformation of BC to ontologies, based on a domain ontology. This method is similar to ontologies alignment methods based on targeted complementary resources, also called background ontologies or support ontologies [18] [19] and [2]. In our proposal, the domain ontology of the IS to design and from which BC to integrate are extracted, plays the role of targeted complementary resource and thus will be our support ontology.

To illustrate our model, we suppose to have two information systems S1 and S2 designed according components based approach; S1 and S2 are to integrate semantically, in order to have a new information system designed from S1 and S2 components.

S1 has a set ($SBC^{s1}$) of $BC^{s1}_1$ .... $BC^{s1}_n$ and S2 has a set ($SBC^{s2}$) of $BC^{s2}_1$ ........ $BC^{s2}_p$. We note $SBC^{S1S2}$ the set representing the union of $SBC^{s1}$ $SBC^{s2}$. So that, all elements of $SBC^{S1S2}$ are candidates for integration. We suppose that the tasks of identification, search, selection of BC and the transformation of BC, assumed to be described in UML, to ontologies ($O_{n+p}$), are made. This transformation is possible [5],[6]. So we have a set of ontologies (SOBC) ($OBC_1$ ... ... $OBC_{n+p}$) produced from $SBC^{S1S2}$.

The semantic integration of business components is proceeding on the following steps:

1. Consider $SBC^{S1S2}$ the set of business components that are candidates for integration.

2. The elements of $SBC^{S1S2}$ are transformed into ontologies, so we obtain a new set of ontologies ($SOBC^{S1S2}$)
3. The ontology alignment technique is applied to the elements of $SOBC^{S1S2}$ .
4. Detection and resolution of semantic conflicts among elements of $SOBC^{S1S2}$ are assumed by the similarity measuring method.
5. Creation of a new Ontology named $OBC_R$. $OBC_R$ describes semantic relations among concepts of $SOBC^{S1S2}$ elements. $OBC_R$ encapsulates knowledge contained in all Business Components which have been integrated.

The diagram below shows our BC integration generic model.

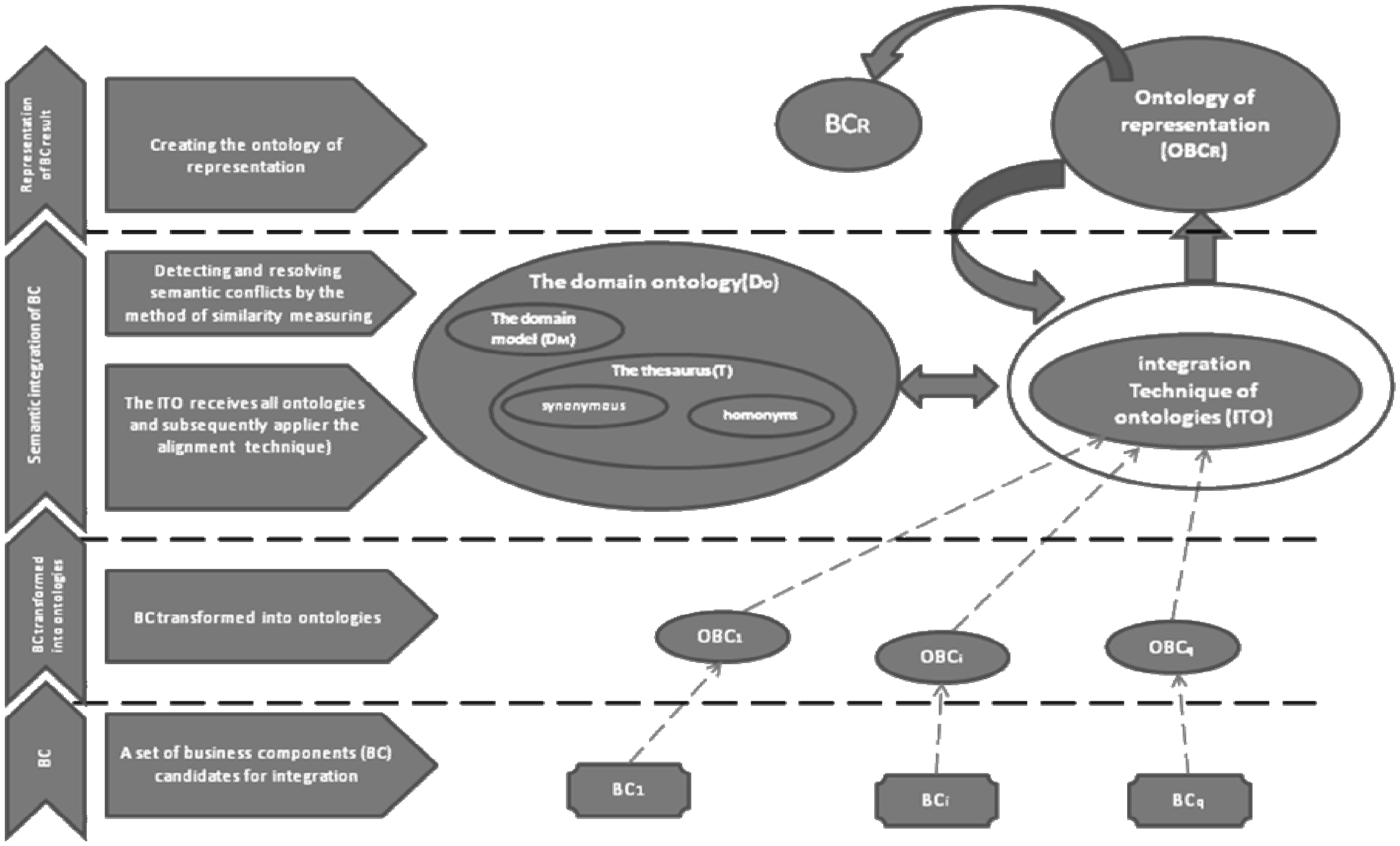

**Figure 4: The model of BC semantic integration.**

The model for BC integration contains several elements:

**A domain ontology:** It models the knowledge in the field of information system, subject of integration. We will reuse vertical domain ontologies. According to [17]: A domain ontology $O_D$ is composed of two elements, the domain model $M_D$ and the thesaurus (T) and will be noted as following $O_D = (T, M_D)$

**An integration technique of ontologies (ITO)** : in this part of the model we achieve treatments on ontologies ($OBC_1$ ... $OBC_q$) produced from the BC. There are several techniques for integrating ontologies: transformation techniques that can deal with a single ontology, techniques of 'mapping' and 'fusion', designed to treat only two ontologies. We used none of these two techniques, considering that we are in a multi-ontologies environment, the only integration that seems appropriate for our case is the ontology alignment technique. This last allows for multiple types of correspondence among concepts (one to one, one to many, many to one and many to many).

**Ontology of representation:** the ontology alignment result, noted ($OBC_R$),it gives us:

- A new Business Component $BC_R$ ,described for example in UML. This BC can help designers and architects in designing a new IS.
- A new ontology $OBC_R$, it can become a source of another treatment or another integration iteration.

### 5.3. Measuring the syntactic similarity.

In this section we propose a method of measurement of syntactic similarity among concepts, this method will be used by the semantic similarity measuring method developed in the next section.

Let be σ ' method of calculating the similarity *Sc* is the set of concepts ∀ *Ci* Є *Sc*, *Ci* is defined by the couple (Tri, $E_{DI}$), Tri: The term referring to the concept *Ci* and $E_{DI}$ all its definitions.

*σ ': Sc × Sc → {0, 1}*
let be $C_1$, $C_2$ concepts in $S_C$ ,Two cases to distinguish:
**Case 1:** $C_1$ and $C_2$ are atomic concepts.
**if** Tr1 =Tr2 **then** σ '($C_1$, $C_2$) = 1
**else** σ '($C_1$, $C_2$) = 0
**end if**
**Case 2:** $C_1$ and $C_2$ are composites.
$C_1$ and $C_2$ written then $C_1 = (C_{11}.., C_{1i}, ... .., C_{1N})$ and $C_2 = (C_{21} ...., C_{2J}, ...., C_{2N})$
the method σ’ is determined as following: σ '(C1, C2) = 1 / n (Σij σ' (C1i, C2J)) 1 <= i, j <= n

When the concepts are syntactically identical, the method σ ' takes the value 1, and 0 otherwise.

### 5.4. Measuring the semantic similarity.

The method of measuring semantic similarity among concepts is based on domain ontology and the method of measuring syntactic similarity σ' defined above.

σ notes the method determining the semantic similarity among concepts, σ is defined as follows:

*σ: Sc × Sc → {0, 1},*
Let be $C_1$ and $C_2$ are two concepts of Sc, $O_D$ domain ontology and T thesaurus.
**Case 1:** C1 and C2 belong to OD and C1 and C2 have a derived term in T:
**If** the derived term is synonymous
**Then** σ (C1, C2) = 1
**Else if** the derived term is homonymous
**then** σ (C1, C2) = 0
**end i**f
**end if**

***Case 2:*** *($C_1$ and $C_2$ belong to OD and $C_1$ and $C_2$ do not have a derived term in T) or ($C_1$ and $C_2$ do not belong to $O_D$):*
*σ ($C_1$, $C_2$) = σ '($C_1$, $C_2$)*
The method σ returns the value 1 when the concepts are synonymous, and the value 0 when they are homonymous. The semantic conflicts among BC are thus detected and resolved. The following table summarizes the different cases that can occur among two concepts C1 and C2.

<table>
<tr><td rowspan="3"></td><td colspan="2">Semantic equality between C1 and C2</td><td colspan="2" rowspan="2">Syntactic equality between C1 and C2.</td></tr>
<tr><td colspan="2">C1 and C2 belong to $O_D$</td></tr>
<tr><td>A derived term synonyms exists in T between C1 and C2.</td><td>A derived term homonyms exists in the T between C1 and C2.</td><td colspan="2">C1 and C2 don't belong to $O_D$</td></tr>
<tr><td>Values of similarity measure</td><td>$\sigma(C1$ , C2)=1</td><td>$\sigma($C1 , C2)=0</td><td>$\sigma'($C1 , C2)=1</td><td>$\sigma'($C1 , C2)=0</td></tr>
<tr><td></td><td>C1 and C2 are synonymous</td><td>C1 and C2 are homonymous</td><td>C1 and C2 are equal</td><td>C1 and C2 are different</td></tr>
</table>

**Table 1: the values of possible measures syntactic and semantic**

## 5.5. Application examples.

We propose in the following three examples to explain and validate our solution.

**Exemple 1:**

Let be BC1 and BC2 two synonymous components belonging to two systems S1 and S2, our method will proceed as follows:

- Determination of ontologies ($OBC_1$ and $OBC_2$) from the two components.

**Case 1: Synonymous components.**

- Calculation of similarity is given as follows: as BC1 and BC2 are synonymous, there will be for each couple of elements (epBC1) from BC1 and (eqBC2) BC2 with their respective ontological concepts (epOBC1) (eqOBC2), σ (epBC1, eqBC2) = 1.
- The overall similarity will be then calculated as follows:

  σ (OBC1, OBC2) = Σij σ (eiBC1, ejBC2)) / n $1 \leq i \leq n$, $1 \leq j \leq n$
  So σ (BC1, BC2) = (n / n) = 1;

We deduce according the method that BC1 and BC2 are synonymous. This confirms our initial hypothesis.

**Case 2: Homonyms components**.

- Calculation of similarity is given as follows: as BC1 and BC2 are homonyms, there must be at least two elements (epBC1) from BC1 and (eqBC2) BC2 with their respective ontological concepts (epOBC1) (eqOBC2), as epOBC1 that is different from eqOBC2 and therefore σ (epBC1, eqBC2) = 0. The overall similarity will be calculated as follows: σ (OBC1, OBC2) = (σ (epBC1, eqBC2) + Σij σ (eiBC1, ejBC2)) / n with ($1 \leq i, j \leq n$ and $(i, j) \neq (p, q)$) So σ (OBC1, OBC2) = (0 + Σij σ (eiBC1, ejBC2) / n) ≠ 1;

We can conclude, according to the method that BC1 and BC2 are homonymous; this confirms our hypothesis.

**Example 2:**

Suppose that we have two business components: the first noted BC1: client (name, age) and the second BC2: client (name, first name). The two components have the same term used to designate client = client; they have the same appearance and therefore σ' (BC1, BC2) = 1; If the concepts associated with the two components belong to the domain ontology and the thesaurus contains a derived term from these two concepts, they will be homonyms and the method of similarity σ will return the value 0. In this case, we can conclude that we have a naming conflict type since we have two concepts having the same appearance and their value similarity (σ) is 0. In case where these concepts do not belong to the domain ontology, we will check their sub-concepts: name, first name and age.

Since these concepts are atomic then σ (name, name) = σ '(name, name) = 1 and σ (first name, age) = σ' (first name, age) = 0. Therefore σ (BC1, BC2) = ½ (σ (name, name) + σ (name, age)) and (σ (name, name) = σ '(name, name), σ (first name, age) = σ' (first name, age)) = ½ (1 + 0) ≠ 1, so σ (BC1, BC2) = 0. These two concepts have the same appearance but with a semantic similarity value equals 0, consequently these concepts have a naming conflict type.

**Example 3:**

This example will illustrate the method of semantic conflicts resolution through the case of merging two libraries. Figures 5 and 6 are representations of UML class diagrams of each of the two libraries. We assume to have the results of the transformation ontologies components of both libraries

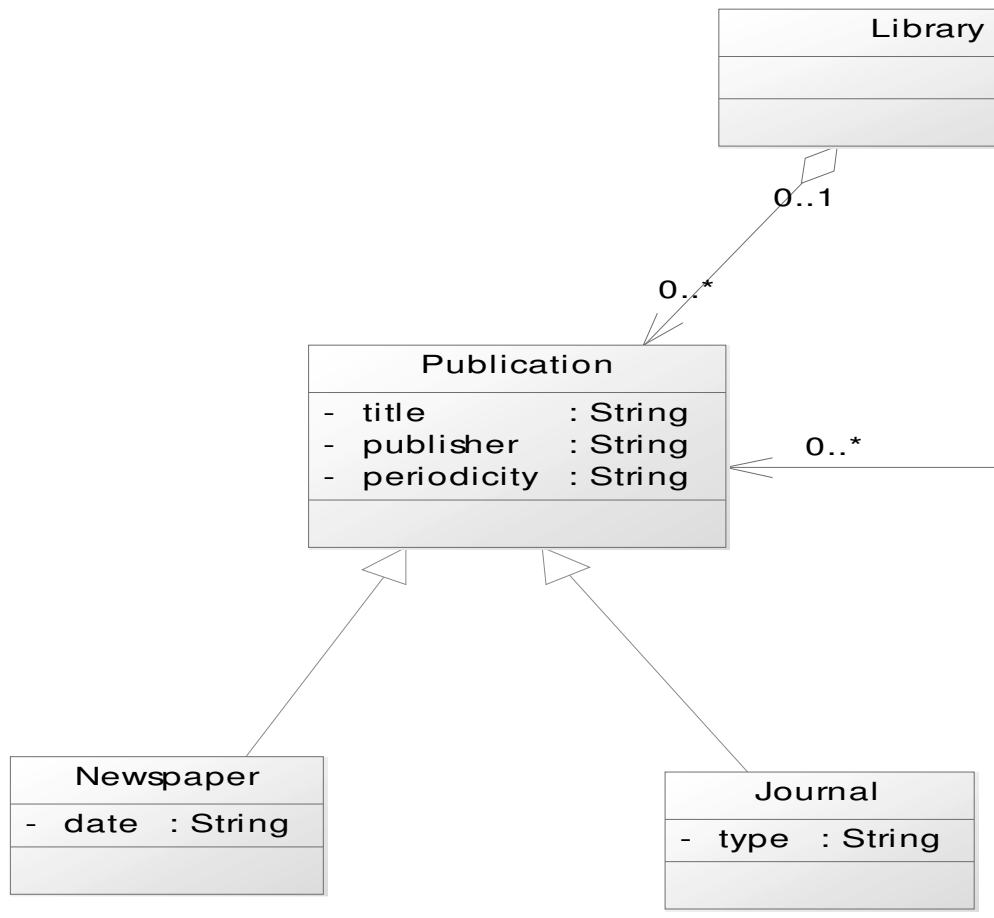

**Figure 5: Class Diagram of library 1**

The library shown in Figure 5 includes newspapers and journals. It deals with accessing online publications. Thus a person can view and read an online publication. A publication is described by its title, publisher and periodicity. A publication may be a newspaper or a magazine.

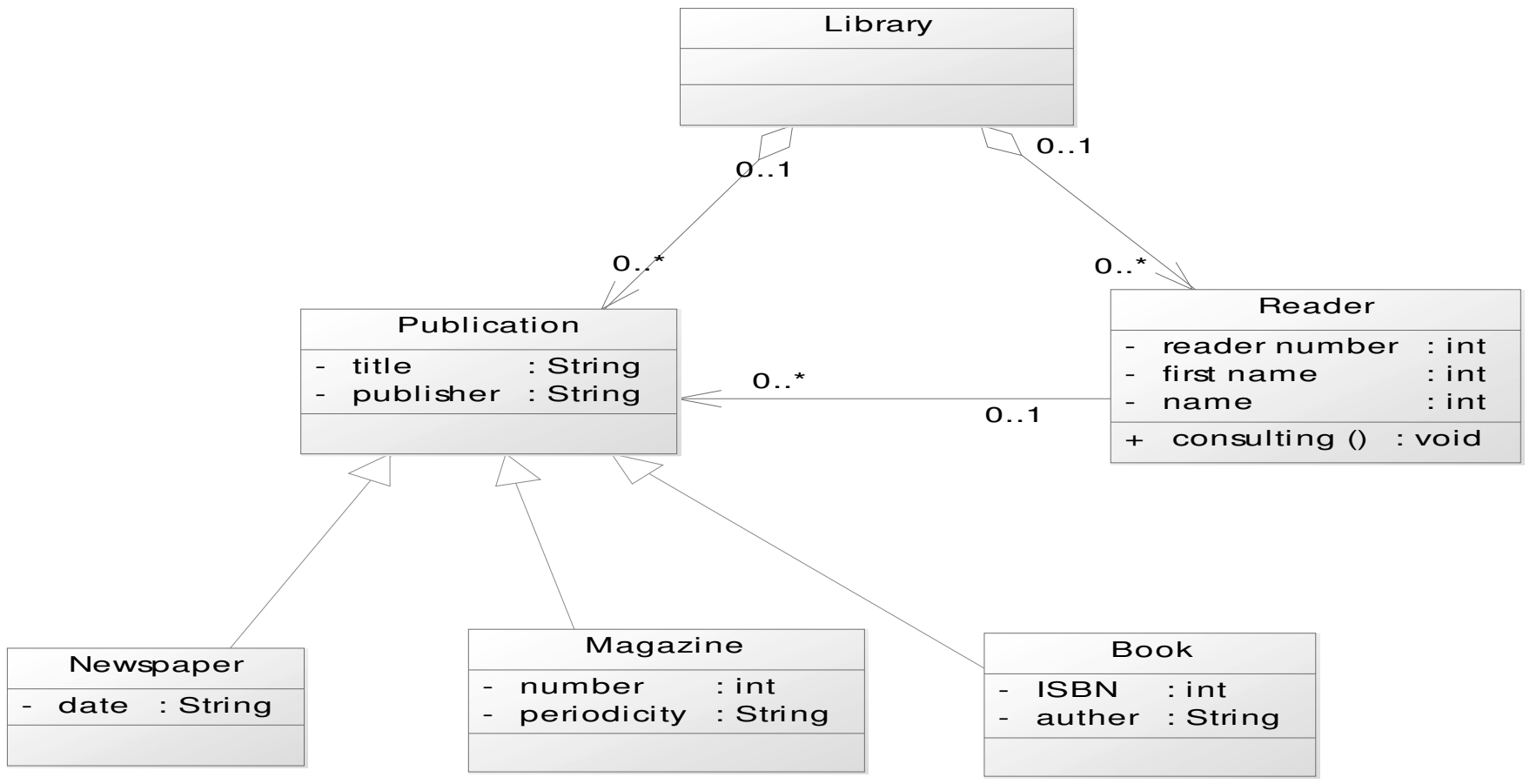

**Figure 6: Class Diagram of library 2**

The library shown in Figure 6 includes newspapers, books and magazines. It also deals with accessing to online publications. Thus, a reader can view and read online publications. A publication is described by its title and publisher. A publication may be a newspaper which is represented by its release date or a book described by its ISBN code and its author or a magazine described by its number and periodicity.

Each class diagram is associated with a diagram of components (Figure 7.8).

Figure 7 shows the diagram component of library 1, this last contain two BC of entity type: "Person" and "publication". The first interface provides the "reading ()" required by the second.

The schart component of library 2, figure 8, contains also two BC of entity type: "Reader" and "publication". The interface provides the first "consulting ()" required by the second.

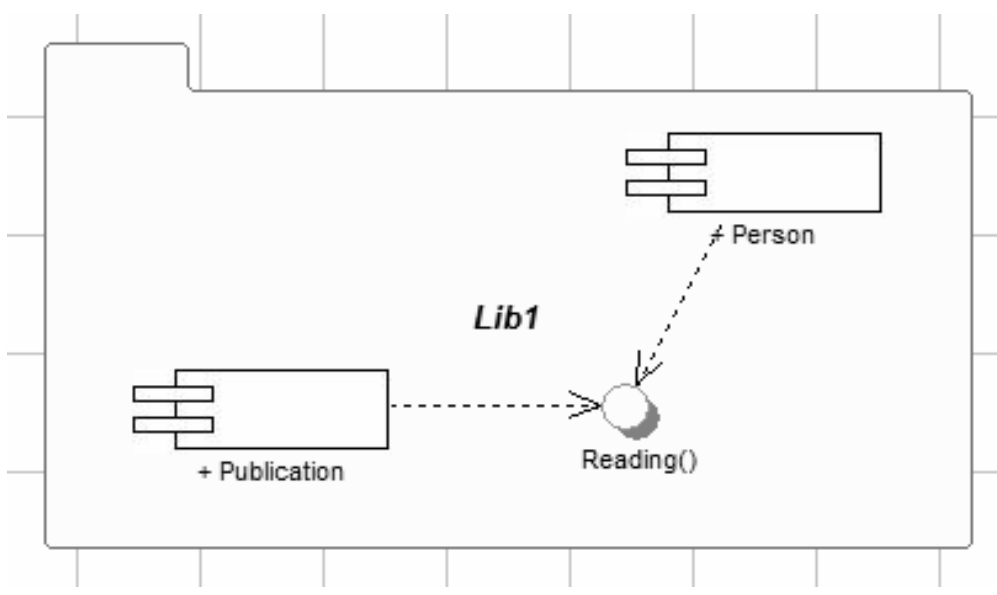

**Figure 7: Component diagram of library 1**

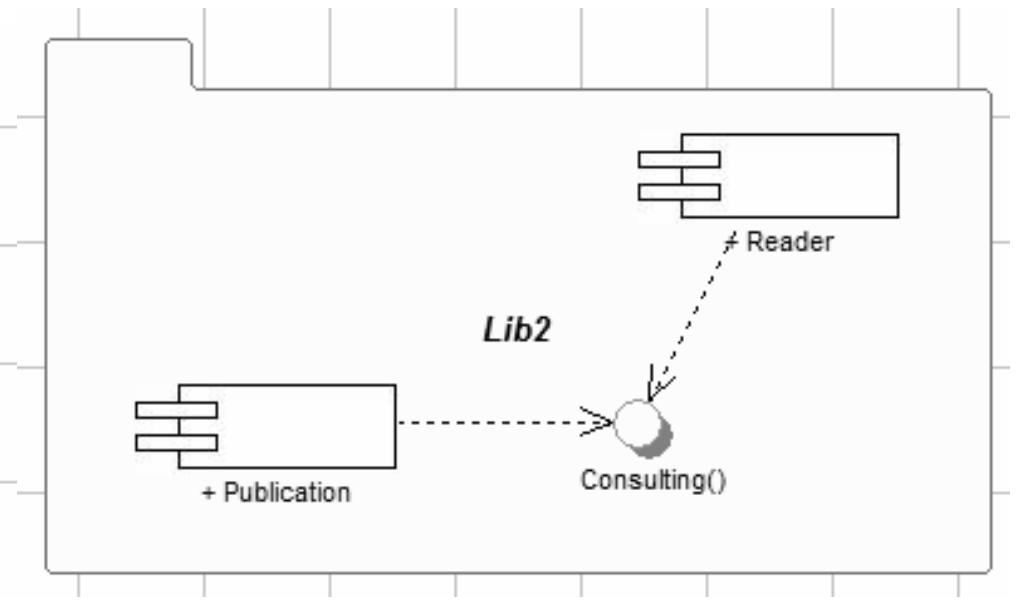

**Figure 8: Component diagram of library 2**

The semantic integration of the four BC will be as following:

1. Identification of business components "Person" and "publication" of "Lib1 and the business components" Reader "and" publication "of Lib2.

2. Production of ontologies corresponding to business components: OntoPerson, OntoPublication, OntoReader and OntoPublication.

3. Reception by the environment that implements TIO, of four input ontologies.

4. Calculating the similarity measure applied to ontologies and their sub-concepts σ (OntoPerson, OntoReader) and σ (OntoPublication, OntoReader).

Table 2 below details the calculation of similarity measures among two ontologies concepts: OntoPerson and OntoReader

| OntoPerson / OntoReader | reader number. | First name | Name | reading() |
|---|---|---|---|---|
| reader number. | 1 | 0 | 0 | 0 |
| First name | 0 | 1 | 0 | 0 |
| Name | 0 | 0 | 1 | 0 |
| consulting () | 0 | 0 | 0 | 1 |

**Table 2: Table of calculating similarity among the concepts**

We will rely on the domain ontology to observe that consulting () and read () are synonymous and therefore: σ (consulting (), reading ()) = 1.

σ (OntoPerson, OntoReader) = ¼ (1 +1 +1 +1 + Σ0) = 1

Ontologies OntoPublication of Lib1 and OntoPublication of Lib2 have the same appearance but have a similarity measure value σ equals 0. We can deduce that there is homonymy and risk of a naming type conflict.

Conclusion: OntoPerson and OntoReader are synonymous; OntoPublication of Lib1 and OntoPublication of Lib 2 are homonyms.

5. - Marking of correspondence among the concepts to solve the problem of semantic conflicts ticks.
   - Creating ontologies of representation.
   - Determining of BC result which represents the knowledge of starting components.

At the end of this process, we obtain as a result a new business component that encapsulates the business knowledge of other components. This new business component can then be used by designers to develop a new information system.

## 6. CONCLUSION.

Our proposal aims is to allow designers and analysts to detect naming semantic conflict type among conceptual business components which are candidate for reuse in an information system project. It consists of a model for semantic integration of business components, and a measuring similarity method detecting and resolving naming conflict type. Our solution is an application of ontologies in the field of BC integration. BC components represented in our work the resources to describe and to integrate. The scope of our solution concerns the conceptual business components available in both centralized and distributed environments.

Examples allowed us to check resolution, we think continue this work by a formal validation of the solution, and then by the search of possibilities of extending it to solve other types of semantic conflicts such measuring and confusion conflicts.

**Authors**

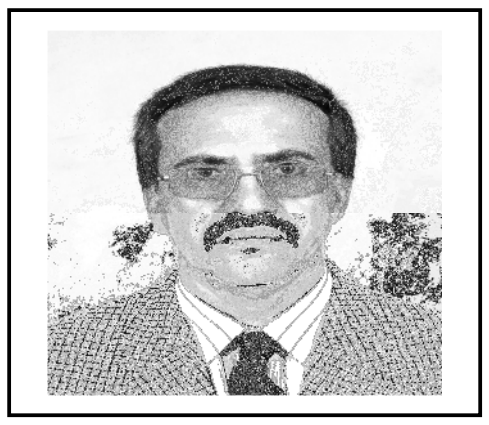

**Larbi Kzaz** received a D.Sc. in Computer Science Domain from Lille University of Science and Technology, France, in 1985.

He does research on computer architecture and information systems. Presently, he is a Professor in Computer Science at ISCAE Business School, Casablanca, Morocco.

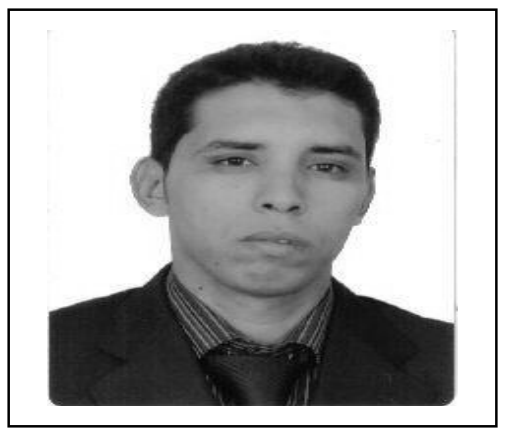

**Hicham ELASRI** is currently a doctoral student in Faculty of Science, University Hassan II Aïn Chock Morocco.

His primary research interest is in semantic interoperability of distributed information system

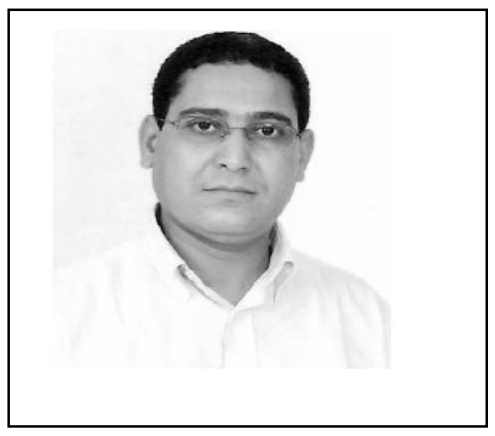

**Abderrahim Sekkaki** received a D.Sc. in Network Management domain from the Paul Sabatier University, France, in 1991: and a Dr. of State Degree from Hassan II University, Morocco, in 2002.

He does research on distributed systems and policies based network management. Presently, he is a Professor in Computer Science at the Hassan II University, Casablanca, Morocco